# Towards Adaptive AI Governance: Comparative Insights from the U.S., EU, and Asia

Vikram Kulothungan*, Deepti Gupta†
*Capitol Technology University, 11301 Springfield Rd, Laurel, Maryland 20708, USA
†Dept. of Computer Information Systems, Texas A&M University - Central Texas, 1001 Leadership Pl, Texas 76549, USA
*vikramk1986@gmail.com, †d.gupta@tamuct.edu

*Abstract*—Artificial intelligence (AI) trends vary significantly across global regions, shaping the trajectory of innovation, regulation, and societal impact. This variation influences how different regions approach AI development, balancing technological progress with ethical and regulatory considerations. This study conducts a comparative analysis of AI trends in the United States (US), the European Union (EU), and Asia, focusing on three key dimensions: generative AI, ethical oversight, and industrial applications. The US prioritizes market-driven innovation with minimal regulatory constraints, the EU enforces a precautionary risk-based framework emphasizing ethical safeguards, and Asia employs state-guided AI strategies that balance rapid deployment with regulatory oversight. Although these approaches reflect different economic models and policy priorities, their divergence poses challenges to international collaboration, regulatory harmonization, and the development of global AI standards. To address these challenges, this paper synthesizes regional strengths to propose an adaptive AI governance framework that integrates risk-tiered oversight, innovation accelerators, and strategic alignment mechanisms. By bridging governance gaps, this study offers actionable insights for fostering responsible AI development while ensuring a balance between technological progress, ethical imperatives, and regulatory coherence.

## I. INTRODUCTION

Artificial intelligence (AI) has emerged as a transformative force in the 21st century, reshaping industries, governance structures, and societal interactions at an unprecedented pace. From generative AI creating human-like text and images to autonomous systems revolutionizing healthcare, finance, and manufacturing, AI's influence is profound and far-reaching. However, this rapid advancement has also introduced complex challenges, including ethical dilemmas, regulatory gaps, and geopolitical tensions. As AI adoption accelerates, regional disparities in AI governance have become increasingly evident, reflecting diverse economic priorities, policy frameworks, and ethical considerations [1]. These disparities not only shape AI's development trajectory but also impact global AI standardization, security, and equitable access. Understanding these variations is essential for balancing technological progress with regulatory coherence while ensuring responsible AI deployment across different political and economic systems.

The absence of cohesive global AI governance has already led to significant real-world consequences. For instance, algorithmic bias in AI-driven hiring systems and facial recognition technologies has raised concerns about systemic discrimination. In 2018, Amazon scrapped its AI-powered hiring tool after discovering that it systematically discriminated against women by favoring male candidates, as it was trained on historical hiring data reflecting past biases. Similarly, facial recognition systems have demonstrated disproportionately higher error rates when identifying people of color, leading to wrongful arrests and privacy violations. Additionally, lack of standardized AI regulations has fueled misinformation and cybersecurity risks. Generative AI tools capable of producing hyper-realistic deepfake videos have been exploited for political manipulation and financial fraud. In 2023, deepfake videos of public figures were circulated online, influencing public opinion and financial markets, raising concerns about election security and media integrity [2]. Without clear AI governance mechanisms, these risks may escalate, threatening democratic processes and economic stability.

Another critical challenge stems from geopolitical competition in AI development. The global race for AI supremacy, particularly between the U.S., EU and Asia, has led to fragmented regulatory landscapes and nationalistic AI policies. While the U.S. fosters a market-driven AI ecosystem prioritizing corporate innovation and minimal regulatory constraints, the European Union enforces a precautionary, rights-based regulatory framework focused on data privacy and ethical safeguards. Meanwhile, Asia employs a state-guided AI model, balancing rapid technological deployment with strong government oversight, particularly in areas like surveillance and public security. These divergent approaches pose challenges to international AI standardization, cross-border collaboration, and ethical AI alignment.

The main contributions of this paper are as follows.

- We present the comparative analysis of AI adoption and governance across three major regions: the United States, European Union, and Asia.
- We present a case study of AI-powered autonomous vehicles and mapped with AI governance.
- We propose AI governance frameworks that balance innovation with ethical oversight and presented the implementation roadmap of these frameworks.
- We also discuss future research priorities to develop a global AI governance framework to enhancing the security.

The remainder of this paper is organized as follows. Section II presents the background and literature review. We discuss the methodology and comparative analysis in Section III and Section IV respectively. Section V presents the case study on autonomous vehicle deployment to identify missing AI governance. The proposed framework is presented in Section VI. Future work and conclusion are discussed in Section VII and Section VIII respectively.

## II. BACKGROUND AND RELATED WORK

Recent scholarship on AI governance highlights the intricate relationship among technological advancement, regulatory mechanisms, and sociopolitical priorities. As AI continues to evolve, different regions have adopted distinct governance models to balance innovation with ethical considerations. In this literature survey, we present an academic perspective on global AI governance frameworks, regional regulatory divergence, and persistent research gaps in the field.

### A. Theoritical Foundations of AI Governance

AI governance is commonly framed through three overlapping paradigms, each reflecting different regulatory philosophies and stakeholder dynamics: Hybrid Governance, Polycentric Governance, and Socio-Technical Systems. The hybrid governance model integrates the state oversight with private-sector self-regulation, allowing governments to set ethical baselines while enabling industry-led innovation [3]. For instance, national AI strategies that establish guidelines for responsible AI while incentivization market-driven advancements. The polycentric governance, a decentralized model where multiple actors-including governments, corporations, and civil society-collaborate to shape AI norms and policies [4]. This framework is particularly relevant for cross-border AI systems, such as automated decision-making in international finance and cybersecurity. Socio-Technical systems contextualize AI within broader societal infrastructures, emphasizing human rights, sustainability, and equitable technological deployment. Governance under this paradigm extends beyond legal framework to include ethical AI design, transparency mandates, and environmental considerations [5].

### B. Regional Governance Models

EU has emerged as a global leader in precautionary AI governance, exemplified by the AI Act, which classifies AI systems by risk level and imposes stricter oversight on high-risk applications, such as biometric surveillance and critical infrastructure automation. This risk-based framework contrasts with the United States sectoral governance approach, creating transatlantic policy misalignment in areas such as AI safety standards and data privacy enforcement [6]. US prioritizes private-sector leadership in AI development, employing a market-driven regulatory approach that relies on voluntary standards and self-regulation. Policies such as NIST AI Risk Management Framework encourage innovation-first strategies while maintaining minimal regulatory intervention [7]. However, this fragmented governance model has raised concerns regarding accountability gaps, particularly in algorithmic bias mitigation, AI fairness, and labor displacement risks [3]. Asian nations have adopted varied AI governance models, blending government oversight with industry policy incentives [8]. China's "New Generation AI Development Plan" emphasizes state-led AI industrialization, prioritizing technological self-sufficiency and regulatory control to drive national AI advancements. In contrast, Japan and South Korea integrate human-centric design principles into their AI governance frameworks, fostering a balance between innovation and consumer protection. These hybrid models challenge Western assumptions about the role of state intervention in technology governance, demonstrating that both authoritarian and democratic regimes can leverage centralized oversight to manage AI risks effectively.

Alfiani et al. [9] conducted a comparative analysis of AI governance strategies in China and the EU, examining their objectives, sectoral approaches, and beneficiaries while proposing policy recommendations for fostering ethical AI development. Hasan [10] analyzed AI legislation in South Asia, comparing it to global regulatory frameworks, highlighting legal disparities, and proposing strategies for effective AI regulation despite economic and structural challenges. Keith [11] examines Global Partnership on Artificial Intelligence (GPAI)'s limited Southeast Asian participation, comparing AI governance policies and proposing strategies to enhance inclusivity through a focus on human capital development. Hine et al [12] developed a philosophy-of-technology-grounded framework to analyze the fundamental differences between US and Chinese AI policies, using Natural Language Processing to compare policy documents and explore historical influences on their AI strategies. Roberts et al. [13] compared China's and the EU's AI strategies, analyzing their goals, sectoral approaches, and beneficiaries, while proposing policy recommendations to enhance ethical AI governance. Luna et al. [14] presented a comparative analysis to facilitate identification of common ground and distinctions based on coverage of the processes by six different regions. In addition, several security models for protecting smart devices are discussed in [15]–[19]. Despite extensive disclosures of AI policies, several critical gaps persist in the academic literature. One major limitation is the lack of integrated governance analysis, as many studies tend to isolate technical, ethical, or industrial perspectives, failing to capture the complex interactions between regulatory frameworks, innovation ecosystems, and societal impact.

## III. METHODOLOGY

This study employs a systematic comparative analysis to examine AI trends and regulatory frameworks across US, EU, and Asia. This methodology focuses on three pivotal trends-foundation models, ethical AI governance, and industrial AI integration-while analyzing regional approaches to development, regulation, and implementation. Our approach combines qualitative and quantitative methods to ensure comprehensive analysis of these complex technological and regulatory landscapes.

## A. Research Design

The selection of research focus areas emerged from a systematic review of current AI developments and their global impact. Foundation models and their applications were selected as the first area of study due to their transformative impact on AI capabilities and broader societal implications. These technologies have fundamentally altered the AI landscape, introducing new possibilities and challenges that demand careful examination. The second focus area, ethical AI governance, was chosen based on the growing importance of regulatory frameworks and compliance mechanisms in ensuring responsible AI development. This aspect has become increasingly critical as AI systems become more prevalent in decision making processes. The third area, industrial AI integration, represents the practical implementation of AI technologies in manufacturing and supply chains, offering concrete examples of how theoretical advances translate into economic impact.

## B. Data Collection and Sources

The regional focus of our study encompasses the United States, European Union, and Asia, representing distinct approaches to AI development and regulation. The United States exemplifies a market-driven innovation model, while the European Union demonstrates a rights-based regulatory approach. Asian nations, particularly China, Japan, and South Korea, present hybrid state-industrial strategies that offer valuable contrasts for analysis.

## C. Analytical Framework

The analytical framework employs four primary evaluation dimensions to assess regional approaches to AI development and regulation. The first dimension examines regulatory frameworks, analyzing policy comprehensiveness, implementation mechanisms, and enforcement effectiveness. The second dimension focuses on innovation ecosystems, considering R&D investment levels, patent activities, and technology transfer efficiency. The third dimension evaluates implementation metrics, including adoption rates and economic impact, while the fourth dimension addresses societal implications, examining ethical compliance and public perception.

## D. Data Analysis

Our analysis employs a mixed-methods approach combining systematic comparison of regulatory approaches with quantitative analysis of adoption patterns. The systematic comparison examines policy frameworks, implementation strategies, and outcomes across regions, while quantitative analysis focuses on statistical evaluation of adoption metrics and performance indicators. Qualitative assessment methods are used to evaluate policy effectiveness and stakeholder impact through detailed case study examination. Content analysis techniques are applied to policy documents and implementation strategies, enabling thorough cross-regional comparison of outcomes.

## E. Methodology Limitations

The study acknowledges several methodological limitations that impact the research. Data constraints present challenges through regional variations in data availability and inconsistencies in reporting metrics. Temporal limitations arise from the rapid evolution of AI technology and the dynamic nature of regulatory environments. Analytical constraints include regional variations in definitions and the inherent complexity of comparative analysis across diverse technological and regulatory landscapes. To address these limitations, the research employs triangulation of data sources and maintains rigorous analytical standards throughout the investigation. This approach ensures that conclusions drawn from the analysis remain robust despite the acknowledged constraints. The methodology provides a comprehensive framework for understanding regional variations in AI development while maintaining awareness of its limitations and the need for continued refinement of analytical approaches.

## IV. COMPARATIVE ANALYSIS

This section presents findings from a comparative analysis of AI adoption and governance across three major regions: the United States, European Union, and Asia. The analysis examines three key dimensions: (1) generative AI development and deployment, and (2) ethical AI governance frameworks

## A. Generative AI Development and Deployment

The analysis reveals distinct regional approaches to generative AI development and implementation. In US, commercial entities lead development efforts, characterized by rapid deployment cycles and market-driven innovation. Notable examples include large language models from technology companies such as OpenAI and Google, which have achieved widespread adoption across content creation, marketing, and software development sectors. The European Union demonstrates a more measured approach, emphasizing regulatory compliance and ethical considerations. The proposed AI Act has established stringent requirements for transparency and copyright protection, resulting in a slower but more controlled adoption rate compared to other regions [6]. This regulatory framework has created a unique environment where innovation proceeds deliberately, with greater emphasis on risk mitigation and societal impact assessment. In Asia, particularly China, the focus lies on rapid scaling of consumer-facing applications, while Japan and South Korea prioritize ethical design principles in specialized markets. Chinese regulatory frameworks mandate content moderation and misinformation controls, creating a unique balance between innovation and governmental oversight [20]. This has resulted in a distinctive ecosystem where rapid deployment coexists with strong state guidance. Regional adoption rates demonstrate significant variance across industries, as shown in TableI:

## B. Ethical AI Governance

The analysis of governance structures reveals three distinct models across regions. The United States operates predominantly through voluntary frameworks, exemplified by the NIST

| Industry Sector | United States | European Union | Asia |
|---|---|---|---|
| Content Creation | 78% | 45% | 65% |
| Marketing | 82% | 38% | 71% |
| Software Development | 73% | 52% | 68% |

TABLE I: Industry Sector Metrics by Region

AI Risk Management Framework [7]. While this approach facilitates rapid innovation, it has resulted in documented instances of algorithmic bias and privacy concerns. The emphasis on self-regulation has created a dynamic but potentially volatile environment for AI development. The European Union has implemented a mandatory governance model incorporating GDPR and AI Act provisions, establishing comprehensive standards for fairness, accountability, and transparency in AI development and deployment. This approach has created a more structured environment for AI innovation, though with potentially slower deployment cycles. The regulatory framework has become a global benchmark for AI governance, influencing policies beyond the EU's borders. Asian countries have developed heterogeneous approaches to AI governance. China has implemented state-aligned ethical guidelines with strict compliance requirements, while Japan and South Korea have focused on collaborative governance emphasizing human-centric design principles. This regional variation has produced a rich tapestry of governance approaches, as reflected in the TableII.

| Dimension | United States | European Union | Asia |
|---|---|---|---|
| Transparency | Medium | High | Variable |
| Accountability | Medium | High | High |
| Privacy Protection | Medium | High | Medium |

TABLE II: Comparison of Different Regions by Dimension

## V. Case Study: Regional AI Governance in Autonomous Vehicle Deployment

The global race to deploy AI-powered autonomous vehicles (AVs) illustrates the impact of regional governance models on commercial success, ethical outcomes, and public trust. This section examines how these three governance models influence AV implementation, using real world case studies to illustrate policy-driven trade offs and sector-specific challenges.

### A. Case Study: AV Deployment Across Regions

*1) United States: Market-Driven deployment:* In US, AV deployment follows a market-driven model, exemplified by Waymo's (Alphabet) operations, where minimal federal oversight enables rapid innovation. By 2023, Waymo's AV had logged over 7 million miles, reducing accidents by 85 percent compared to human drivers [21]. The absence of federal mandates for algorithmic transparency, however raises ethical concerns regarding decision-making in crash scenarios. Private-sector-led safety certifications replace centralized approvals, accelerating deployment but leaving regulatory gaps in liability and accountability.

| Region | Example | Key Strength | Key Weakness | Governance Model |
|---|---|---|---|---|
| United States | Waymo (Alphabet) | Rapid deployment (7M+ miles logged) | Ethical opacity, lack of oversight | Market-driven, voluntary regulation |
| European Union | Mercedes-Benz Level 4 AVs | High public trust (78% approval) | Slower commercialization | Strict regulatory compliance (AI Act, GDPR) |
| Asia | Baidu Apollo (China) | Large-scale adoption (500K+ robotaxis) | State-controlled AI priorities | Hybrid state-market governance |

TABLE III: Comparison of Autonomous Vehicle Governance Models by Region

*2) European Union: Regulation-First Approach:* In contrast, the European Union enforces a regulation-first approach, prioritizing ethical AI compliance and public trust. Mercedes-Benz Level 4 AVs require compliance with the EU AI Act and GDPR mandates, ensuring transparency through pre-market conformity assessments [6]. These policies have led to higher public approvals of AV technologies but also cause significant commercialization delays, with approval processes extending several months longer than in the US, impacting global competitiveness.

*3) Asia: State-Guided AV Expansion:* Meanwhile, Asia's AV governance follows a state-guided model, where government policies drive large-scale deployment. Baidu Appllo, China's largest AV operator, have deployed Robotaxis across 11 cities, backed by government partnerships [22]. Unlike the US and EU, China mandates state-approved AI training datasets, aligning AV decision-making with national safety priorities. However, strict data localization laws limit foreign firms access, restricting global AV standardization and creating barriers for international interoperability [20].

### B. Key Insights: Governance Models Shaping AV Adoption

AV governance is a balancing act between innovation, regulation, and public trust. Across regions, three competing approaches reveal distinct priorities and trade-offs:

- Speed vs. Safety: The US fast-tracks AV deployment, embracing market driven innovation but lacks standardized ethical oversight. Meanwhile, the EU enforces rigorous safety and transparency mandates, ensuring trust at the cost of slow commercialization. Asia finds a middle ground, leveraging state-backed infrastructure to scale AV adoption quickly, yet raises concerns over centralized AI governance.
- Regulatory Fragmentation: AVs are designed to navigate roads, yet struggle to navigate regulations. EU certified AVs may lack compliance in Arizona, where policies are minimal, while China's data localization laws prevent global AV firms from deploying seamlessly. The lack of harmonized standards restricts cross-border expansion, limiting global AV interoperability.
- Industrial Strategy and Market Power: China's state-driven coordination enables mass AV deployment, the EU prioritizes ethical AI leadership, and the U.S. benefits from private-sector agility. Yet, each model carries risks that overregulation can hinder market growth, while underregulation raises safety concerns.
- The Road Ahead: AV technology is accelerating, but policy remains in the rearview mirror. Bridging regional governance gaps will require adaptive regulatory frameworks, international AI safety benchmarks, and ethical alignment—ensuring that AVs don't just move fast, but move responsibly.

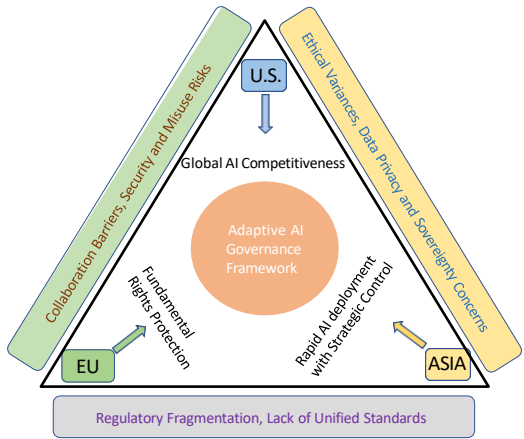

Fig. 1: Global AI Governance Spectrum

## VI. Towards Adaptive AI Governance Frameworks

The analysis of AV governance reveals broader patterns in AI regulation that require adaptive frameworks balancing innovation with ethical oversight [3]. Building on these insights while generalizing beyond transportation applications, the next section focuses on developing cross-sector AI governance principles. The figure 1 presents the proposed global AI governance spectrum, which represents among three different regions.

### A. Hybrid Governance Architectures

A proposed governance framework that draws on regional strengths to create a more adaptive, inclusive, and effective AI regulatory ecosystem:

*1) Risk-Tiered Oversight (Inspired by the EU Model):*
- Implement a tiered regulatory approach that categorizes AI systems based on risk level [6]
- High-risk AI applications in critical sectors (e.g., healthcare, energy, criminal justice) should undergo rigorous auditing, transparency mandates, and ethical impact assessments [23]
- Lower-risk AI applications (e.g., commercial and consumer services) should benefit from streamlined regulatory pathways to encourage innovation while maintaining baseline accountability

*2) Innovation Accelerators (Inspired by the US Model):*
- Establish regulatory sandboxes to support the controlled testing and development of emerging AI technologies such as quantum machine learning and neuro-symbolic AI [24].
- Introduce sunset clauses requiring periodic legislative review (e.g., every three years) to ensure that regulations evolve in tandem with technological advancements.
- Foster public-private collaboration to develop voluntary industry standards that complement formal regulatory frameworks.

*3) Strategic Alignment Mechanisms (Inspired by the Asian Model):*
- Create national AI governance councils with equal representation from industry, academia, and government to ensure balanced decision making in AI policy and research investments [25].
- Align AI development goals with national priorities, such as economic competitiveness, workforce adaptation, and ethical AI deployment.
- Leverage government-backed AI infrastructure and public sector Research and Development initiatives to enhance AI accessibility and technological self-sufficiency [1].

By integrating these regional strengths into a unified governance model, this framework seeks to harmonize innovation driven policies with ethical oversight, ensuring AI's responsible deployment across industries and societies.

### B. Dynamic Regulatory Tools

To ensure AI governance remains responsive to technological advancements and societal risks, this section proposes some examples of dynamic regulatory mechanisms that adapt to real-world AI deployment challenges. These kind of tools emphasize proactive risk mitigation, performance-based oversight, and continuous monitoring to foster responsible AI innovation.

By embedding these adaptive tools into AI governance structures, policymakers can create a regulatory ecosystem that evolves with AI technologies, ensuring safety, accountability, and public trust without stifling innovation.

### C. Key Initiatives for International Alignment

Achieving a balance between international harmonization of AI governance and regional policy autonomy is essential to foster global AI cooperation while respecting local regulatory priorities. This section outlines key initiatives aimed at aligning AI governance frameworks across borders while maintaining national sovereignty in ethical and policy decisions.

| Mechanism | Implementation | Monitoring Metric |
|---|---|---|
| Algorithmic Impact Bonds | Require AI developers to post financial bonds as a safeguard against potential societal harms. If AI systems cause measurable harm (e.g., bias, safety failures), bond payouts fund remediation efforts. | Bond value is proportionate to system complexity, sector sensitivity, and deployment scale. Higher-risk AI applications (e.g., autonomous weapons, predictive policing) require larger bonds. |
| Adaptive Licensing | Implement a tiered licensing framework where AI systems receive conditional market approval, with continued access contingent on real-world safety and fairness performance. | Measured by incident rates per 10 million user interactions, ensuring AI systems demonstrate reliability before full-scale deployment. Failure to meet safety benchmarks results in license suspension or modification. |
| Ethics Stress Testing | Mandate adversarial simulations and red-teaming exercises for high-risk AI applications (e.g., autonomous vehicles, financial AI, medical diagnostics) to identify vulnerabilities before deployment. | Track the number and severity of failure modes detected in pre-deployment testing, ensuring AI systems undergo rigorous evaluation against ethical, security, and safety threats [23]. |

TABLE IV: AI Governance Mechanisms

*1) Mutual Recognition Agreements (MRAs):*

- Facilitate cross-border AI deployment by establishing agreements that recognize equivalent regulatory and safety standards across jurisdictions.
- Ensure AI systems certified in one region can operate in another, provided they meet shared baseline safety and ethical requirements.
- Two core components of MRAs:
  1) Equivalent Certification Processes: Establish common criteria for AI safety and risk assessments, allowing AI models approved under one regulatory framework (e.g., EU AI Act) to be recognized in other markets (e.g., U.S., Asia).
  2) Bilateral Data Adequacy Determinations: Enable cross-border data flows by mutually recognizing data protection regimes, ensuring compliance with privacy standards like GDPR while allowing innovation-friendly data sharing.

*2) Common Technical Protocols:* Develop international benchmarks and adopt ISO standards to ensure AI interoperability, accountability, and sustainability [26]. Key focus areas include:

- AI System Documentation: Standardized model documentation (e.g., model cards, dataset manifests) to enhance transparency and auditability of AI applications [27].
- Incident Reporting Frameworks: Establish a global AI risk and failure reporting system to track and share information on AI-related incidents, ensuring proactive governance responses [6].
- Energy Efficiency Benchmarks: Define sustainability metrics for AI systems, ensuring energy-efficient model training and deployment to reduce environmental impact [26].

By integrating these global-local governance mechanisms, AI policies can achieve a balance between international collaboration and national sovereignty, ensuring responsible AI deployment worldwide.

## VII. FUTURE RESEARCH PRIORITIES

As AI governance evolves, new challenges and opportunities emerge, requiring continued interdisciplinary research. This section highlights key research priorities that can inform future policy development, ensuring AI regulation remains effective, adaptive, and globally relevant.

### A. Cultural Dimensions of AI Trust

This future research will explore how regional cultural values shape public perceptions of AI decision-making, highlighting the influence of societal norms on AI adoption. It examines differences in AI acceptance across various governance models, including the U.S. market-driven approach, the EU's rights-based framework, and Asia's hybrid state-market strategy. Additionally, the study identifies key factors that influence trust in AI systems, such as transparency, human oversight, and societal narratives surrounding automation, providing insights into the conditions necessary for fostering public confidence in AI technologies.

### B. Predictive Governance Tools

This future research will examine the potential of AI-driven policy simulation systems that utilize machine learning to model and anticipate the real-time impacts of regulatory decisions. It explores how AI can enhance regulatory agility by predicting unintended consequences, identifying enforcement gaps, and recommending adaptive policy adjustments to improve governance efficiency. Furthermore, the study critically evaluates the ethical implications of incorporating AI into decision-making processes, emphasizing the necessity of human oversight, accountability, and transparency to mitigate risks and uphold democratic governance. By addressing these aspects, this research aims to provide insights into the responsible integration of AI in policymaking while balancing innovation with ethical considerations.

### C. Economic Trade-Off Metrics

This future study will aim to develop quantitative models that evaluate the trade-offs between rapid AI innovation and precautionary regulatory measures, providing a structured approach to balancing progress with oversight. It investigates the economic impact of various AI governance strategies, analyzing how strict versus flexible regulations influence investment flows, market adoption rates, and global AI competitiveness. Furthermore, the study proposes standardized evaluation frameworks to equip policymakers with data-driven tools for assessing the costs and benefits of AI interventions, ensuring regulatory decisions that foster both technological advancement and responsible governance.

By advancing these research priorities, policymakers and industry leaders can develop more effective AI governance frameworks that align with technological progress, economic imperatives, and societal values.

## VIII. Conclusion

The divergent AI governance models adopted by the U.S., EU, and Asia reflect unique regulatory philosophies, yet their fragmentation poses challenges to global interoperability, ethical coherence, and policy coordination. To bridge these gaps, an adaptive governance framework must integrate risk-tiered oversight, innovation accelerators, and strategic alignment mechanisms, ensuring that AI deployment remains both responsible and dynamic. A key priority is the adoption of agile regulatory tools that evolve with technological advancements. Mechanisms such as algorithmic impact bonds, ethics stress-testing, and modular legislation provide responsive safeguards against emerging AI risks while fostering innovation. International harmonization efforts, including mutual recognition agreements and standardized AI documentation protocols, are essential to facilitating cross-border collaboration.

Future AI governance must embrace predictive regulatory models, leveraging AI itself to anticipate policy gaps and optimize oversight. Research should further explore the intersection of AI ethics, cultural trust dynamics, and economic trade-offs to ensure governance strategies align with societal values and industry needs. By synthesizing regional strengths into a unified yet flexible governance approach, this study lays the groundwork for a resilient, globally coordinated AI regulatory ecosystem that safeguards ethical principles while enabling technological progress.

## IX. Acknowledgment

This work is partially supported by the US National Science Foundation grant 2431531.